\documentclass[aps,prl,reprint,groupedaddress]{revtex4-2}
\usepackage{graphicx,float}
\usepackage{amsmath}
\usepackage{array}
\newcolumntype{P}[1]{>{\centering\arraybackslash}p{#1}}

\begin{document}


\title{Determination of the absorbed dose in the brain with DICOM images using GATE/GEANT4 and MATLAB}


\author{Walter Sabalza-Castillejo}
\email[]{waltersabalza@gmail.com}
\thanks{Corresponding author}
\affiliation{Hospital Universitario de Caracas, Universidad Central de Venezuela, Caracas, Venezuela}

\author{Miguel Martín-Landrove}
\email[]{landrove.martin@ucv.ve}
\affiliation{National Institute for Bioengineering, INABIO, Universidad Central de Venezuela, Caracas, Venezuela}
\affiliation{Centro de Diagnóstico Docente Las Mercedes, Caracas, Venezuela}


\date{\today}

\begin{abstract}
In this study, the absorbed dose calculation for the cerebral region is refined using the Geant4 software, along with the ROOT and GATE programs. This approach provides a detailed statistical analysis of positron trajectories within a water medium, using a voxel-sized source based on PET imaging. By simulating positron displacement within a 4×4×4 $mm^3$ voxel containing radioactive material mixed with water, and bordered by voxels of identical density, the exact absorbed dose in the brain can be precisely determined. PET images, processed in MATLAB, will complement this dosimetric analysis. Results are compared to established guidelines, such as MIRD Pamphlet 17 and ICRP 106, to validate the accuracy and consistency of the absorbed dose calculations. The primary aim of this research is to optimize the absorbed dose in the cerebral region using digital images from the Philips PET-CT Gemini GXL6 scanner at the Radiotherapy and Nuclear Medicine Service, University Hospital of Caracas. The absorbed dose is calculated with the GATE simulator, a Monte Carlo-based tool supported by the Geant4 and ROOT frameworks. The focus is on evaluating the positron displacement from the central voxel to neighboring voxels. Post-simulation, the program’s algorithms provide detailed data on absorbed doses for both gamma and beta radiation emitted by $^{18}F$. This simulation data is then integrated into PET DICOM images within MATLAB. Geometric calculations are applied to further enhance absorbed dose information by leveraging the voxel characteristics of the images, thus optimizing the absorbed dose depiction. Results will be benchmarked against standards like ICRP 109, MIRD Pamphlet 2, and ICRP 53 for validation and optimization.
\end{abstract}


\maketitle

\section{Introduction}
Internal dosimetry is a critical component of nuclear medicine, facilitating the accurate measurement and calculation of radiation doses absorbed by tissues in the body during diagnostic and therapeutic procedures. Precise dose assessment is particularly vital in treatments involving radionuclide therapy, where high-dose regions must be accurately targeted to achieve effective therapeutic outcomes while minimizing exposure to surrounding healthy tissues \cite{Bolch2009}. Recent advances in imaging technology, notably the use of Digital Imaging and Communications in Medicine (DICOM) standards, have allowed improved visualization and analysis of patient anatomy and radiotracer distribution, laying the groundwork for more refined dose calculations.

The use of DICOM images enables the integration of patient-specific anatomical and physiological information into the dose assessment, which is essential for personalized treatment planning \cite{Hindorf2010}. By importing these images into computational models, such as those developed in the GEANT4 Monte Carlo simulation toolkit, researchers and clinicians can simulate radiation transport and interactions with high spatial resolution \cite{Agostinelli2003}. The GATE (Geant4 Application for Tomographic Emission) platform, based on GEANT4, has emerged as an effective tool to simulate positron emission tomography (PET) studies, enabling researchers to model complex imaging systems, optimize acquisition protocols, and evaluate image quality in relation to dosimetry \cite{Jan2004,Jan2011,Sarrut2021}.

In PET dosimetry, GATE has been widely applied to simulate realistic clinical scenarios, including patient-specific dosimetry and the interaction of positron-emitting radio tracers with biological tissues \cite{Parodi2007,Jan2011}. Through these simulations, PET imaging can be optimized for both diagnostic and therapeutic applications, allowing accurate estimation of tracer distribution, absorbed doses, and improved understanding of dose-response relationships \cite{Sarrut2022}. This approach provides enhanced anatomical specificity and dose accuracy, using DICOM's anatomical information and GATE's high-fidelity simulation capabilities.

This study presents a comprehensive approach to internal dosimetry by combining DICOM imaging data with GATE-based simulations tailored to PET applications. Through this integration, our goal is to improve the accuracy of dose estimations in nuclear medicine, highlighting potential applications and validation methods that emphasize patient safety and optimized treatment outcomes in clinical practice.
\section{Materials and methods}.
\subsection{Dose modeling}
The absorbed dose modeling of a voxel begins with simulations in the GATE program. Following these simulations, the results are applied to DICOM brain PET images of a patient injected with 18F-FDG, obtained from Philips Gemini GXL6 equipment at the Hospital Universitario de Caracas. The images are processed to isolate the region of interest (ROI), and then the calculation of the absorbed dose is modeled in MATLAB using both the simulation results of GATE and the intensity values of each voxel.

\begin{figure}[H]
\centering
\includegraphics[width=7cm]{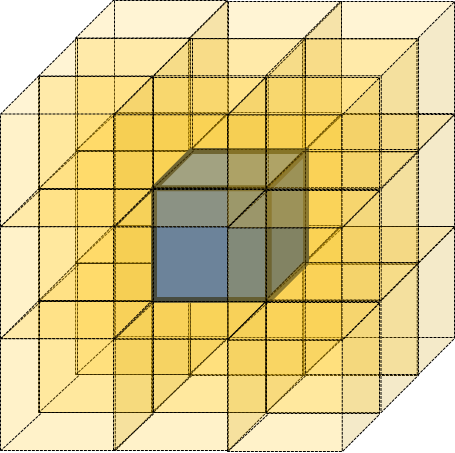}
\caption{Physical simulation model where the central voxel contains radioactive material dissolved in water and the adjacent ones only water}
\label{InterIntra}
\end{figure}

In the GATE simulation, the physical model consists of a 4×4×4 $mm^{3}$ voxel filled with water, containing fluorine-18 uniformly distributed with an initial activity of 1000 Bq. This central voxel is surrounded by adjacent voxels on all sides, as illustrated in Figure \ref{InterIntra}. The absorbed dose results are compared with simulations using smaller voxel sizes of 1×1×1 $mm^{3}$ and 2×2×2 $mm^{3}$ to evaluate the impact of voxel dimensions on dose calculations. In addition, an absorbed dose calculation is performed in a 10 cm water phantom to assess the influence of gamma radiation on distant voxels.

Once the simulation data are obtained, they are integrated into MATLAB. The simulation results provide the number of beta particle interactions occurring both within the central voxel and in the adjacent voxels, including both frontal and diagonal neighbors. In MATLAB, the DICOM images are loaded and sectioned to isolate the relevant areas. The absorbed dose is then calculated by assessing the percentage of interactions in the central voxel and adjacent voxels, applying the absorbed dose equations from MIRD Pamphlet 17 \cite{Pamphlet17}.

\subsection{Monte Carlo method.}

The transport of particles and their interactions can be modeled as a stochastic process, well-suited for Monte Carlo simulations. These simulations are widely used in applications such as high-energy physics and medicine, as they detail each particle’s trajectory and interactions individually \cite{Cantone2011}.

\textbf{Phase Space}: Each dimension in phase space corresponds to a degree of freedom of the particle. Three dimensions represent its real space position ($x, y, z$), while other dimensions represent momentum components ($p_x, p_y, p_z$), energy ($E$), and angular parameters (e.g., $\Theta$ and $\phi$). Additional dimensions account for properties such as quantum numbers, spin, particle count, and others. Each particle is represented as a point in phase space, which varies with time \cite{Cantone2011}.

The fundamental quantity describing a group of particles is the phase space density, denoted as $n(t,x,y,z,p_x,p_y,p_z)$. This density represents the number of particles within an infinitesimal volume element in phase space. The quantity $\Psi = n \vec{v}$ is known as the \textit{angular flux}, a general radiometric quantity. Angular flux $\Psi$ is defined as the derivative of \textit{fluence} $\Phi$ with respect to all phase space coordinates: time, energy, and solid angle (or direction vector).

\begin{equation}
\Psi = \frac{\partial \Phi}{\partial t \partial E \partial \vec{\Omega}} = \dot{\Phi}_{E\vec{\Omega}}.
\end{equation}

Angular flux $\Psi$ is a differential quantity; however, in Monte Carlo simulations, it is often integrated over one or more phase space dimensions, such as spatial coordinates, energy, angles, or fluence.

\begin{equation}
\Psi = \int_E \int_{\vec{\Omega}} \int_t \vec{\Omega}_{E\vec{\Omega}} d E d \vec{\Omega} dt.
\end{equation}

In Monte Carlo calculations, the time dependence is often not explicitly considered, and the differential fluence with respect to energy, $\Phi(E) = \frac{d\Phi}{dE}$, becomes the primary quantity of interest \cite{Cantone2011}.

\textbf{Boltzmann Space:} At any point in phase space, the change in particle density $n$ within an infinitesimal phase space volume is equal to the sum of all "production terms" minus all "destruction terms." This balance equation can be expressed in terms of the angular flux $\Psi$.

\begin{widetext}
\begin{equation}
\frac{1}{v}\frac{\partial \Psi (x)}{\partial t} + \vec{\Omega} \cdot \nabla \Psi (x) + \sum_t \Psi(x) - S(x) =  \int_{\Omega} \int_E \Psi(x) \sum_s (x' \leftarrow x) dx'\\
\end{equation}
\end{widetext}
where $x$ represent all space of phase coordinates $\vec{r}, \vec{\Omega}, E, t$.\\

The other elements of the Boltzmann equation have the following physical meanings:

\begin{itemize}
\item The term $\frac{1}{v} \frac{\partial \Psi(x)}{\partial t}$ represents the change in angular flux over time and accounts for the energy decay of particles.
\item $\vec{\Omega}\cdot\nabla \Psi(x)$ describes the change in angular flux due to translational movement, with no change in energy or direction.
\item $\sum_t \Psi(x)$, where $\sum_t$ is the macroscopic total cross-section (inverse of the mean free path), represents an absorption term.
\item $S(x)$ denotes the particle source.
\item The double integral involving $\sum_S$, the macroscopic scattering cross-section, represents the change in angular flux due to directional changes without changes in the particle’s position.
\end{itemize}

All Monte Carlo particle transport calculations aim to solve the Boltzmann equation in its integral form, that is, by integrating over all possible particle states.

\subsubsection{Analog Particle Transport (Photons).}

The probability that a photon will interact with the homogeneous medium at a distance $s$ from the surface typically follows an exponential attenuation law, which can be written as:
\begin{equation}
p(s)ds= \mu(E)\exp^{-\mu (E) s} ds
\label{Aten}
\end{equation}
where $\mu$ is the linear attenuation coefficient of the medium for the photons. To find the mean length $<s>$  at which interactions occur, we can calculate the expected value of $s$ based on the probability density function, assuming that the medium is infinitely extended under the surface:
\begin{equation}
<s>=\int_0^{\infty} s p(s) ds = \mu(E) \int_0^{\infty} s \exp^{-\mu(E)s} ds = \frac{1}{\mu(E)}.
\end{equation}

To create a dimensionless parameter that characterizes a distance $s$ relative to the mean interaction length, we can define the adimensional parameter $\lambda$ as:

\begin{equation}
\lambda=\frac{s}{<s>}=\mu (E) s
\end{equation}
and (\ref{Aten}) can be written as,
\begin{equation}
p(\lambda) d\lambda = \exp^{-\lambda} d\lambda.
\label{atenLaw}
\end{equation}

Using (\ref{atenLaw}), the cumulative distribution function is given by,
\begin{equation}
P(\lambda)=\int_0^{\lambda} d \lambda' p(\lambda') = \int_0^{\lambda} d \lambda' exp^{-\lambda'} = 1 - \exp^{-\lambda}
\end{equation}

This function grows monotonically in $[0 , \infty)$. To determine the distance $\lambda_1$ to the first interaction site using the transformation method, a uniform random number $\xi_1$ is generated from the interval $[0,1]$ and:
\begin{equation}
\xi_1 = 1 - \exp^{-\lambda_1} \rightarrow \lambda_1 = -\ln(1-\xi_1).
\end{equation}

Depending on the photon energy, four processes are possible: photoelectric, Rayleigh scattering, Compton scattering and pair production. Each is represented by an interaction coefficient as a parameter at the interaction site:
\begin{equation}
\mu_{tot}(E)=\mu_{Ph}(E)+\mu_{Rs}(E)+\mu_{Cs}(E)+\mu_{Pair}(E)
\label{CoefFot}
\end{equation}

The interval $[0,1]$ is divided into four parts.

\begin{equation}
\begin{matrix}
[P_0,P_1] & : & \text{Photoelectric} \\
[P_1,P_2] & : & \text{Rayleigh Scattering} \\
[P_2,P_3] & : & \text{Compton Scattering} \\
[P_3,P_4] & : & \text{Pair Production}
\end{matrix}
\end{equation}
with:
\begin{widetext}
\begin{equation}
P_0=0, P_1=P_0+\dfrac{\mu_{Ph}}{\mu_{tot}}, P_2=P_1+\dfrac{\mu_{Rs}}{\mu_{tot}}, P_3=P_2+\dfrac{\mu_{Cs}}{\mu_{tot}}, P_4=1.
\end{equation}
\end{widetext}

The type of interaction is determined by using a second random number $\xi_2$ within the interval $[0,1]$. Once the interaction type is identified, the characteristics of secondary particles, such as their energies and scattering angles, are determined by sampling from the probability distributions defined by the corresponding differential cross sections \cite{Hubbell1999,Hendee2013}.

\textbf{Photon interaction coefficient}
To calculate the material parameter from the total cross-section $\sigma_i(E)$ of the i-th element in the material, the general formula typically involves summing contributions from all constituent elements, weighted by their respective atomic or molar fractions. The calculation is expressed as:

\begin{equation}
\mu(E)=\sum_i N_i(\vec{r})\sigma_i(E)).
\end{equation}
where $N(\vec{r})$ is the number of atoms of the element per unit volume at a point $\vec{r}$ and is calculated as:

\begin{equation}
N_i(\vec{r})=\frac{\rho(\vec{r})\omega_i(\vec{r})}{m_uA_i(\vec{r})}=\frac{\rho(\vec{r}) N_A \omega_i(\vec{r})}{M_i(\vec{r})}
\end{equation}
where $\rho(\vec{r})$ is the mass density, $\omega_i(\vec{r})$ is the weight fraction of the i-th element, $A_i(\vec{r})$ is the relative atomic mass of the i-th element, $M_i(\vec{r})$ is the molar mass of the i-th element, $m_u= 1.6605388x10^{-27}$ kg is the unit of mass, and $N_A$ is the Avogadro constant.

The interaction probabilities $\mu_k(E,\rho)$ $(k=Ph,Rs,Cs,Pair)$ from equation (\ref{CoefFot}) can be determined directly by the following approximation:

\begin{equation} \label{StopPowPhot}
\mu_k(E,\rho)=\frac{\rho}{\rho^{W}} f_k(\rho)\mu_k^{W}(E)
\end{equation}
$\mu_k^{W}(E)$ are the interaction coefficients for water and $\rho^{W}$ is the water density \cite{Hendee2013}.

\subsubsection{Charged Particle Transport.}
\begin{figure}[h!]
\centering
\includegraphics[width=\columnwidth]{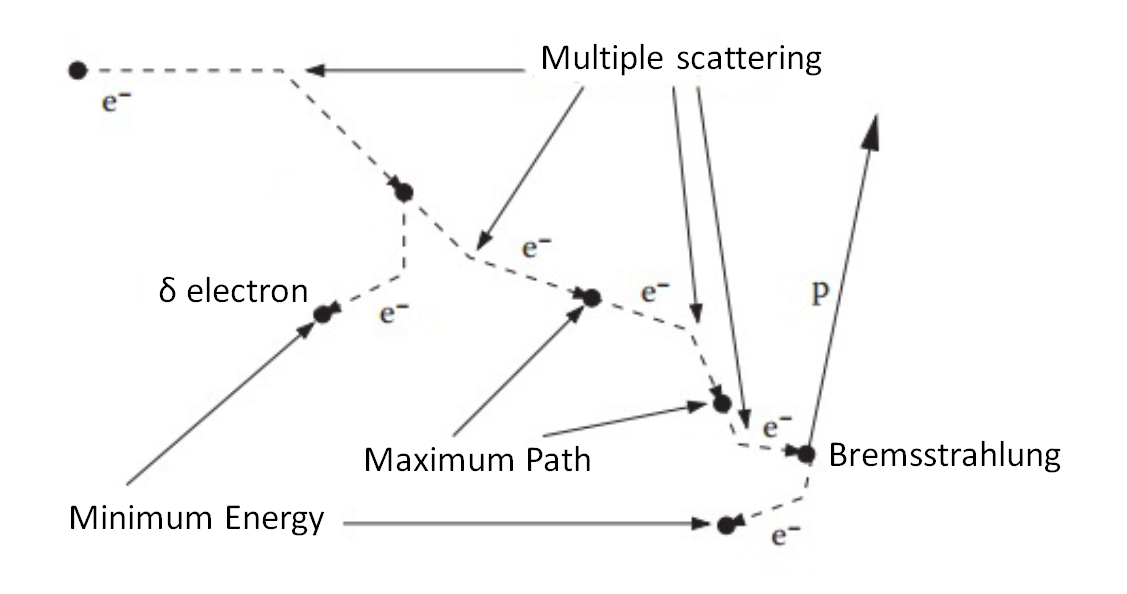}
\caption{Physical model of the movement of $e^-$.}
\label{CH2}
\end{figure}

The Continuous Slowing Down Approximation (CSDA) \cite{Halbleib1980} and condensed-history (CH) \cite{Berger1963,Larsen1992,Kawrakow1998} models are essential for simulating charged particle interactions efficiently. These methods simplify the detailed simulation of the numerous small-angle scatterings and energy losses (Figure \ref{CH2}) that charged particles undergo when passing through matter. 

\begin{figure}[h!]
\centering
\includegraphics[width=\columnwidth]{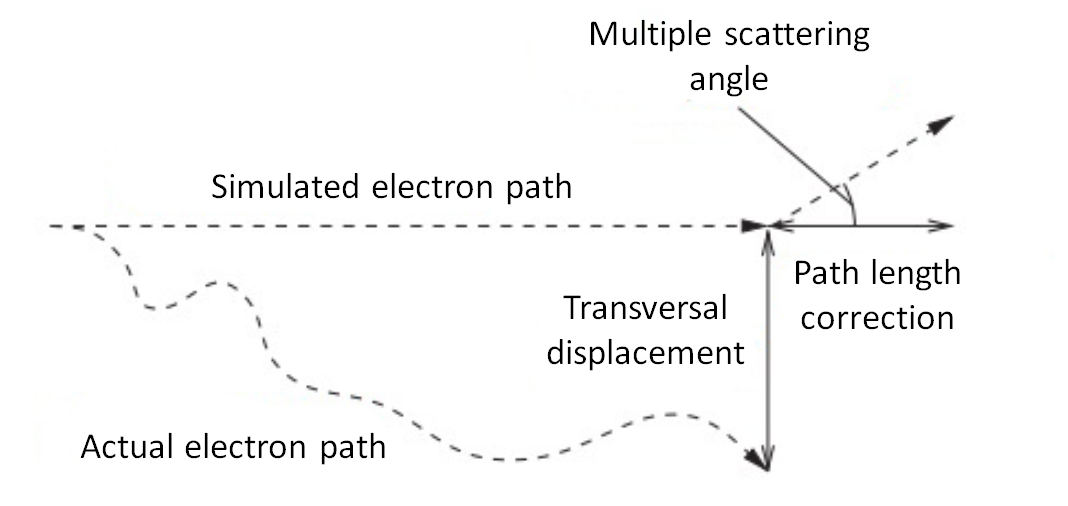}
\caption{CH model of the motion of a charged particle, in this case an $e^-$.}
\label{CH1}
\end{figure}

The CH technique, introduced by Berger in 1963, revolutionized charged particle simulation by balancing computational efficiency and physical accuracy, making it a cornerstone of modern particle transport simulations. Figure \ref{CH1} visually contrast the detailed interaction paths and their simplified representations under the CH approximation.

\textbf{Continuous energy loss.} The continuous energy loss in the CH model is calculated using the linear stopping power, which represents the average energy lost per unit path length by a charged particle as it traverses a material. The energy loss $dE$ over an infinitesimal path length $ds$ is given by the linear stopping power \cite{Berger1984},

\begin{equation}
\small
L(\vec{r}, E, E_c, k_c) \equiv -\left( \frac{dE}{ds} \right)_{res} = L_{col}(\vec{r}, E, E_c ) + L_{rad}(\vec{r}, E, k_c)
\label{LossEnergy}
\end{equation}
with

\begin{widetext}
\begin{equation}
L_{col}(\vec{r}, E, E_c) \equiv - \left( \frac{dE}{ds} \right)_{res,col}, \\
L_{rad}(\vec {r}, E, k_c) \equiv - \left( \frac{dE}{ds} \right)_{res,rad}
\end{equation}
\end{widetext}

Additionally it can be rewritten as follows:

\begin{equation}
\begin{matrix}
L_{col}(\vec{r}, E, E_c)= N(\vec{r})\int_0^{E_c} dE' E' \sigma _{col}(\vec{r}, E, E' ) \\
L_{rad}(\vec{r}, E, k_c)= N(\vec{r})\int_0^{k_c} dk' k' \sigma _{col}(\vec{r}, E, k' ).
\end{matrix}
\label{StopPow}
\end{equation}
where $N(\vec{r})$ is the number of scattered targets per unit volume at point $\vec{r}$. The step length $s$ for an electron with initial energy $E_0$ that loses energy $\Delta E$ due to CH transport is determined by integrating the energy-loss equation (\ref{LossEnergy}):

\begin{equation}
s=-\int_{E_0}^{E_1} \dfrac{dE}{L(\vec{r}, E, E_c, k_c)}=\int_{E_1}^{E_0} \dfrac{dE}{L (\vec{r}, E, E_c, k_c)},
\end{equation}

with $E_1=E_0-\Delta E$ \cite{Hendee2013}.

\subsection{Creation of the Model in GATE.}
The simulation elements are placed in this program by starting the interface with the definition of the world (\textbf{World}) in which all the physical processes will occur. The world will be defined by the user and in this case a space of 20 cm on each side is defined in the x, y, and z axes. This volume must not be smaller than any other volume involved in the simulation and no object must be outside the World environment.

\begin{verbatim}
/gate/world/geometry/setXLength 20.0 cm
/gate/world/geometry/setYLength 20.0 cm
/gate/world/geometry/setZLength 20.0 cm
\end{verbatim}

The world must have a materials library that the user must include, and this is predetermined by the database file named \textit{GateMaterials.db}, this database is provided by the GATE software and defines a wide range of materials , molecules, atoms, gases and air, without this file the system does not compile.

\begin{small}
\begin{verbatim}
/gate/geometry/setMaterialDatabase GateMaterials.db
\end{verbatim}
\end{small}

Then the other elements are created to provide physical information of the system and an arbitrary name is placed, in this case the material that will be used will be named \textit{VoxelPET} and the type of elements, whether it is a geometric figure, a phantom, a voxel, among other materials, in this case the type of material is a Voxel:

\begin{small}
\begin{verbatim}
/gate/world/daughters/name      VoxelPET
/gate/world/daughters/insert    ficticiousVoxelMap

\end{verbatim}
\end{small}

\begin{figure}[H]
\centering
\includegraphics[width=\columnwidth]{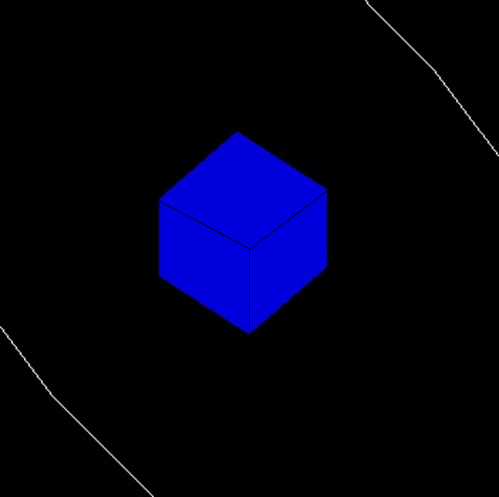}
\caption{Cubic system where the central voxel and the adjacent voxels are located, all of them are blue}
\label{Model}
\end{figure}

The radioactive material, shape, characteristics and activity are then introduced into the system. In addition, the physical events of the same will be placed so that the system takes them into account.

\begin{widetext}
\begin{verbatim}
/gate/source/addSource                          F18VolSource
/gate/source/F18VolSource/setActivity           1000. becquerel
/gate/source/F18VolSource/gps/particle          e+
/gate/source/F18VolSource/setForcedUnstableFlag true
/gate/source/F18VolSource/setForcedHalfLife     6586.2 s
/gate/source/F18VolSource/gps/energytype        Fluor18
/gate/source/F18VolSource/gps/type              Volume
/gate/source/F18VolSource/gps/shape             Para
/gate/source/F18VolSource/gps/halfx             2.0 mm
/gate/source/F18VolSource/gps/halfy             2.0 mm
/gate/source/F18VolSource/gps/halfy             2.0 mm
/gate/source/F18VolSource/gps/angtype           iso
/gate/source/F18VolSource/gps/centre            0.0 0.0 0.0 cm
/gate/source/F18VolSource/visualize             red
\end{verbatim}

\begin{verbatim}
/gate/physics/addProcess PhotoElectric
/gate/physics/addProcess Compton
/gate/physics/addProcess GammaConversion
/gate/physics/addProcess LowEnergyRayleighScattering

/gate/physics/addProcess ElectronIonisation
/gate/physics/addProcess Bremsstrahlung
/gate/physics/addProcess PositronAnnihilation

/gate/physics/addProcess MultipleScattering e+
/gate/physics/addProcess MultipleScattering e-

/gate/physics/processList Enabled
/gate/physics/processList Initialized
\end{verbatim}
\end{widetext}

The JamesRandom simulation engine \cite{Marsaglia1990,Sepehri2020} will also be introduced as it demonstrates greater precision in the results for low interactions. The simulation is set to acquire data at intervals of approximately 1 second, yielding 10 dose measurements for each run. These measurements are then averaged to obtain the final dosimetric values before proceeding with the full simulation.

\begin{verbatim}
/gate/application/setTimeSlice   1 s
/gate/application/setTimeStart   0 s
/gate/application/setTimeStop    10 s
\end{verbatim}

\begin{verbatim}
/gate/random/setEngineName    JamesRandom
/gate/random/setEngineSeed    default
/gate/random/verbose          1
\end{verbatim}

\begin{figure}[!h]
\centering
\includegraphics[width=\columnwidth]{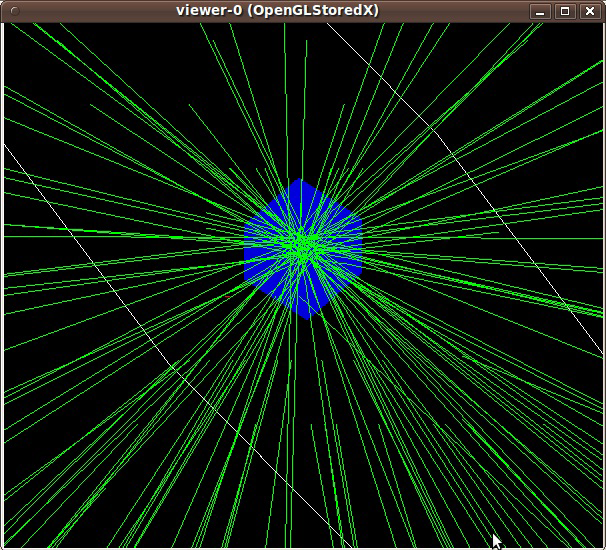}
\caption{Simulation of the interaction of $^{18}F$ in GATE}
\end{figure}

\subsubsection{Obtaining the Data.}
The absorbed dose rate $\dot{D}$ is given by

\begin{equation}
\dot{D}=\frac{kA \sum _{i} y_{i} E_{i} \phi _{i}}{m},
\label{DA}
\end{equation}

where:
\begin{itemize}
    \item $\dot{D}$ is the absorbed dose rate (in rad/h or Gy/s),
    \item A is the activity (in mCi or MBq),
    \item y is the number of radiations with energy E emitted per nuclear transition,
    \item E is the energy per radiation (in MeV),
    \item $\phi$ is the fraction of emitted energy absorbed in the target,
    \item m is the mass of the target region (in g or Kg),
    \item k is a proportional constant (in rad.g/$\mu$Ci.h.MeV or Gy.Kg/MBq.s.MeV).
\end{itemize}

To determine the total internal dose for a specific organ, the time-dependent component of the activity must be integrated over time to obtain the cumulative dose \cite{Stabin2008}.

The cumulative dose equation can be expressed as:

\begin{equation}
D=\frac{k \tilde{A} \sum _{i} y_{i} E_{i} \phi _{i}}{m} ,
\label{DAT}
\end{equation}

This formula allows the calculation of the cumulative dose, $D$, by incorporating the total accumulated activity, $\tilde{A}$, over time.

\begin{widetext}
\begin{equation}
\tilde{A}= \int_{0}^{\infty} A(t) dt= \int_{0}^{\infty} (f A_{0}) e^{- \lambda_{e} t} dt= \\ \frac{fA_{0}}{\lambda _{e}}=1.443f A_{0}T_{e},
\label{ActAcum}
\end{equation}
\end{widetext}
where $A_{0}$ represents the administered activity, $f$ denotes the fraction of activity directed to the region of interest, and ($f .A_{0}$) is the initial amount of activity within that region. The effective half-life ($T_{e}$) is a crucial parameter for calculating cumulative activity and cumulative dose \cite{Stabin2008}.

The data from the GATE program simulation provide the absorbed dose per second for both the central and adjacent voxels. To determine the total absorbed dose, the dose per second result is used. By applying equation \eqref{DA}, we obtain the function for the absorbed dose rate. However, because this dose rate intrinsically depends on activity, equations \eqref{DAT} and \eqref{ActAcum} are used to calculate the total dose. Since the absorbed dose obtained is an intrinsic value, it will serve as the primary variable for the final dose calculation.

\begin{equation}
D=1.443T_eD(t),
\label{DTAMC}
\end{equation}
where the effective lifetime, $T_{e}$, is derived using the biological half-life, $T_{b}$ for fluorodeoxyglucose (FDG) elimination from the body, which is approximately 6 hours. The relationhip for $T_{e}$ is given by:

\begin{widetext}
\begin{equation}
T_{e}=\frac{T_{p}T_{b}}{T_{p}+T_{b}}=\frac{109 min \times 360 min}{109 min+360 min}=84.12 min=5047.2 sec.
\label{Teff}
\end{equation}
\end{widetext}

So \eqref{DTAMC} looks like this:

\begin{equation}
D=1.443 * 5047 sec * D(t)
\label{DTOTAL}
\end{equation}

The absorbed dose values are related to both the calculated activity and the activity derived from DICOM images. To accurately determine the source activity, a calibration adjustment is performed using a cylindrical phantom during the SUV (Standardized Uptake Value) calibration procedure. For this purpose, a 2 mCi phantom, uniformly distributed, is introduced to ensure accurate activity measurements across the imaging system.

\subsection{Introduction of dicom images using the MATLAB program.}

The PET DICOM images are selected with tomographic correction applied for a full-body scan. The brain region is isolated using the MATLAB command \textit{roicolor}, which restricts the selection to this area. Each image slice is then combined into a 3D volume using the \textit{cat} command. Calibration of the data is performed using a cylindrical phantom to establish the SUV (Standardized Uptake Value) calibration factor.

The absorbed dose results from GATE are provided in sieverts (Sv) for a water medium with dimensions of 12mm x 12mm x 12mm, within which a central 4mm x 4mm x 4mm region contains fluorine-18 (18F) with an activity of 1000 Bq over 1 second. This simulation is repeated 10 times to determine the average dose per second. The results are then imported into MATLAB, where the absorbed dose is calculated voxel-by-voxel based on the DICOM images.

In MATLAB, the voxel-wise absorbed dose calculation uses the average absorbed dose values for both the central and adjacent voxels, as per Equation \eqref{DTOTAL}. The absorbed dose values from GATE are converted by dividing by 1000 to obtain per-activity values and are then incorporated into the MATLAB code alongside the SUV calibration factor.

\subsection{Representation of results.}
At the end of the simulation, several files containing the system's produced iterations are generated according to the specified conditions. A key file in this study is \textbf{dose.raw}, which is opened and analyzed using the ImageJ program. By examining its histogram, we obtain the results for the absorbed doses in Gy, as shown in Table 2.1 of \cite{OGate2019}.

The results indicate that the dose in the central area is between 87.62 \% and 90.77 \% of the total calculated dose. In adjacent voxels, the dose does not exceed 2.8 \% to 3.78 \% of the total dose calculated for the entire volume. The standard deviation of the central voxel across the ten simulations is 1.45 \%. The computation time for each iteration on the system ranges between 4 and 5 hours.
The average of the total dose within the volume is calculated and the results are shown in Table \ref{PromCent}, with a standard deviation of 1.22 \%.

\begin{figure}[h!]
\centering
\includegraphics[width=\columnwidth]{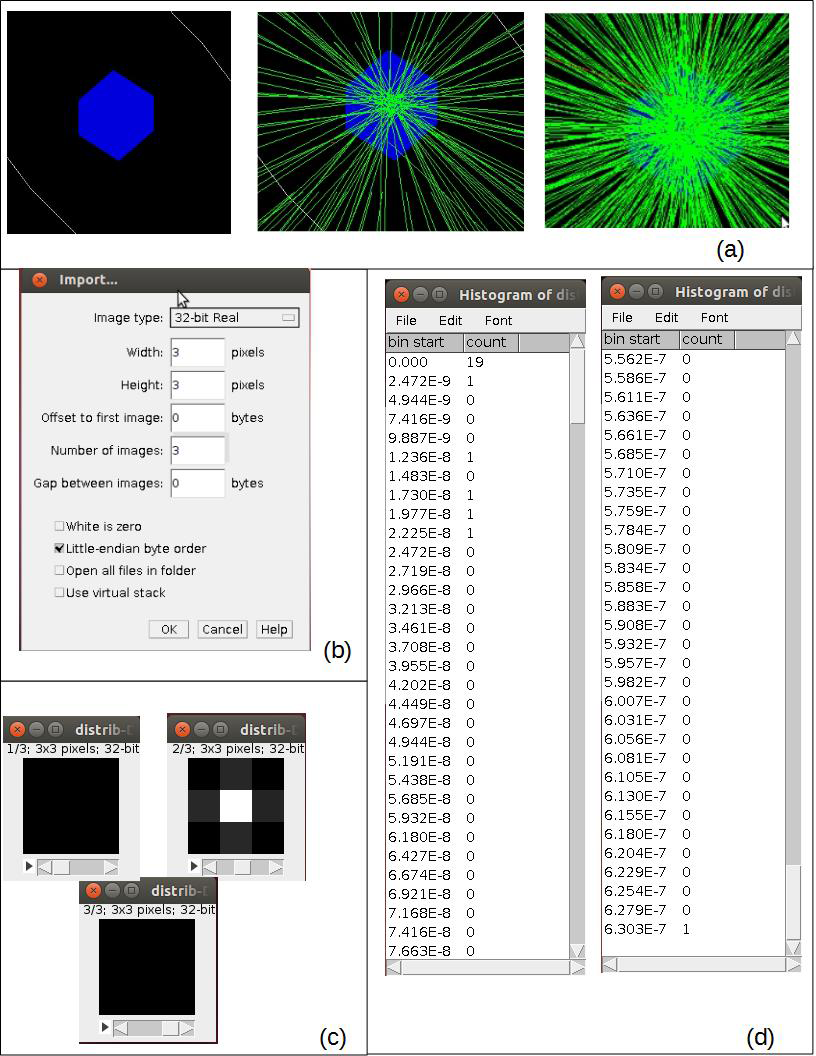}
\caption{Steps in Gate and ImageJ. (a) Iterations are observed at the start, initial stages of data collection; middle, progress halfway through the calculation and at 95\% completion, highlighting the convergence of the absorbed dose values. The data are saved in a .raw file for visualization and analysis in ImageJ. (b) ImageJ window for parameters of the .raw to be introduced. (c) Dose intensity maps. The clearest (brightest) areas correspond to the highest absorbed dose, typically at the center of the 3x3x3 voxel grid. (d) Once the center of the voxel grid is selected, histograms for the absorbed dose are obtained}
\label{IterFin}
\end{figure}

\begin{figure}[h!]
\centering
\includegraphics[width=\columnwidth]{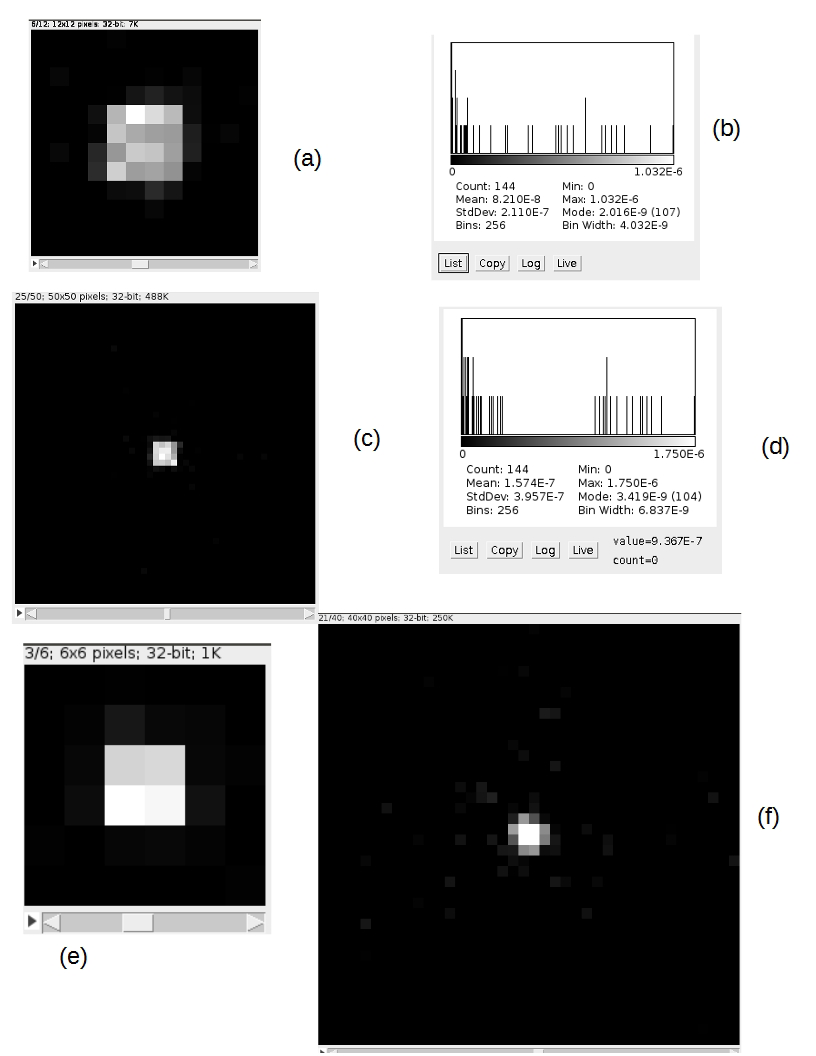}
\caption{Results of absorbed dose in the central area using 1x1x1 $mm^3$ resolution (a), while in (b) the histogram of this result is observed, the same thing also happens in (c) for a volume of 12x12x12 $mm^ 3$ and (d) for a volume of 40x40x40 $mm^3$ for a resolution of 1x1x1 $mm^3$. In (e) and (f) the same results are obtained but using the resolution of 2x2x2 $mm^3$}
\label{DoseGate}
\end{figure}

\begin{figure}[h!]
\centering
\includegraphics[width=\columnwidth]{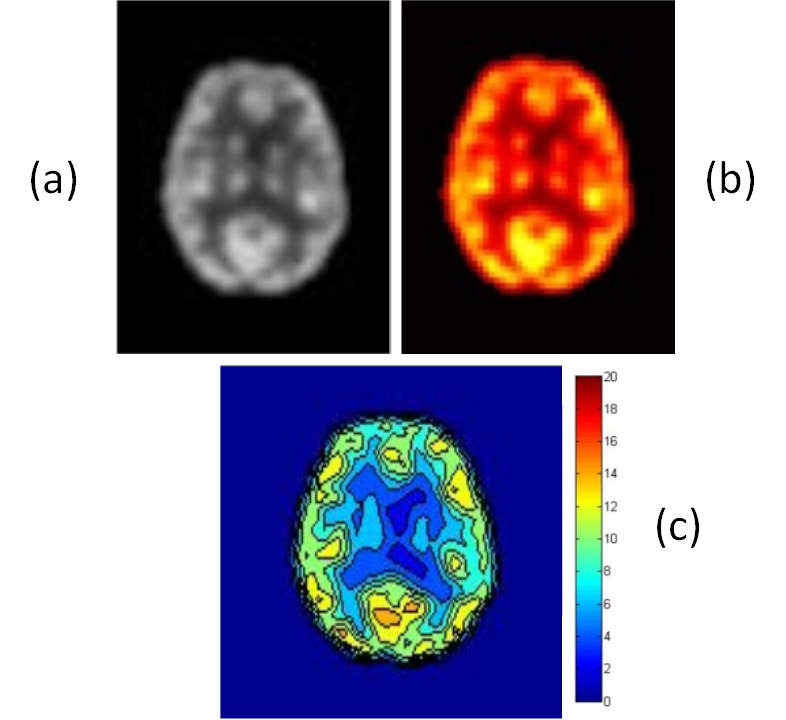}
\caption{Absorbed Dose in the brain area using images in MATLAB. (a) PET image corrected by attenuation, (b) the result of the simulated total absorbed dose and (c) isodose map  represented in color levels, the blue color corresponds to the lowest dose in the calculation, and the red color to the highest dose value.}
\label{DoseDicom}
\end{figure}

In MATLAB, the absorbed dose is calculated using the values from Table \ref{PromCent}. The steps in GATE to obtain the absorbed dose values are detailed in Figure \ref{IterFin}. Finally, absorbed dose images from the GATE simulator are displayed in Figure \ref{DoseGate}; the isodose map of a PET tomographic slice is shown in Figure \ref{DoseDicom}.

\section{Results and Discussion}
One of the great advantages of this calculation is that the absorbed dose is measured per voxel, providing detailed spatial resolution. This is a key improvement over methods like those in ICRP 106 or anthropomorphic phantoms used by MIRD, which estimate the dose across entire organs. This precision allows for localized analysis of dose distribution, which is particularly relevant for heterogeneous tissues or treatments like targeted radiotherapy. The GATE program is highlighted for its reliability, contingent on accurately inputted physical parameters. Using the JamesRandom simulation engine, dosimetric values deviate by less than 2\% when 10 simulations are performed. This consistency underscores the robustness of the simulation framework and its independence from voxel size. The absorbed dose at the center of the radioactive material represents the 80-95 \% of the total dose within the volume.

\begin{table} [!h]
\centering
\caption{Average absorbed dose.}
\small
\begin{tabular}{|P{2cm}|P{2.5cm}|P{2.5cm}|}
\hline
\textbf{Location} & \textbf{Average Absorbed Dose (Gy)} & \textbf{Standard Deviation (Gy)} \\ 
\hline
\textbf{Central voxel} & $6.33x10^{-7}$ & $9.24x10^{-9}$ (1,45 $\%$)\\
\hline
\textbf{All voxels} & $7.10x10^{-7}$ & $8.72x10^{-9}$ (1,22 $\%$)\\
\hline
\end{tabular}
\label{PromCent}
\end{table}

\begin{table} [!h]
\centering
\caption{Results from a 40x40x40 $mm^3$ with a 2x2x2 $mm^3$ voxel phantom, this result demonstrates that the gamma dose is within the calculation along with the positive beta.}
\begin{tabular}{|P{2.5cm}|P{2.5cm}|P{2.5cm}|}
\hline
\textbf{Absorbed Dose (Gy)} & \textbf{Number of Voxels} & \textbf{Voxel Position} \\ \hline
0 & 1564 & Away \\
$2.98x10^{-9}$ & 14 & Away\\
$5.96x10^{-9}$ & 6 & Away\\
$8.94x10^{-9}$ & 5 & Away\\
$1.79x10^{-8}$ & 2 & Adjacent \\
$2.09x10^{-8}$ & 1 & Adjacent \\
$3.58x10^{-8}$ & 1 & Adjacent  \\
$3.87x10^{-8}$ & 2 & Adjacent  \\
$5.66x10^{-8}$ & 1 & Adjacent  \\
$5.93x10^{-7}$ & 1 & Central \\
$5.96x10^{-7}$ & 1 & Central \\
$6.61x10^{-7}$ & 1 & Central \\
$7.60x10^{-7}$ & 1 & Central \\\hline
\end{tabular}
\label{DosisGde}
\end{table}

Regarding the obtained results, the absorbed dose vary slightly by region, with an average dose ranging between 6-12 mGy depending on the specific area. In the parietal, occipital, and frontal lobes, the average absorbed dose is approximately 6-10 mGy. However, as we approach the cerebellum, the absorbed dose increases to 16 mGy. This regional variation demonstrates the utility of voxel-based approaches in capturing localized dose differences. The ICRP 106, Addendum 3 \cite{ICRP106} reports a whole-brain dose between 10–30 mGy for 10 mCi administered activity. Similarly, the MIRD Pamphlet 19 \cite{Hays2002} suggests an organ-wide brain dose of ~14–20 mGy. Therefore, The study's results (6–16 mGy by region) align closely with these reference ranges, validating the method's credibility and accuracy. This voxel-based approach not only aligns well with established dosimetric data but also enhances the granularity of absorbed dose estimation, making it a valuable tool for personalized radiological assessments and treatment planning.

\begin{table} [!h]
\centering
\caption{Results from a 2 x 2 x 2 $mm^3$ voxel of 12x12x12 $mm^3$ phantom.}
\begin{tabular}{|P{2.5cm}|P{2.5cm}|P{2.5cm}|}
\hline
\textbf{Absorbed Dose (Gy)} & \textbf{Number of Voxels} & \textbf{Voxel Position} \\ \hline
0 & 17 & Away \\
$2.68x10^{-9}$ & 2 & Away\\
$5.36x10^{-9}$ & 3 & Away \\
$8.04x10^{-9}$ & 1 & Away \\
$1.07x10^{-8}$ & 1 & Adjacent \\
$1.61x10^{-8}$ & 1 & Adjacent \\
$1.88x10^{-8}$ & 1 & Adjacent \\
$2.14x10^{-8}$ & 1 & Adjacent \\
$2.68x10^{-8}$ & 1 & Adjacent \\
$4.02x10^{-8}$ & 1 & Adjacent \\
$4.82x10^{-8}$ & 1 & Adjacent \\
$5.36x10^{-8}$ & 2 & Adjacent \\
$5.81x10^{-7}$ & 1 & Central \\
$6.08x10^{-7}$ & 1 & Central \\
$6.56x10^{-7}$ & 1 & Central \\
$6.83x10^{-7}$ & 1 & Central \\\hline
\end{tabular}
\label{DosisFantoma888}
\end{table}

\begin{table} [!h]
\centering
\caption{Results from a 12x12x12 $mm^3$ 1x1x1 $mm^3$ voxel phantom.}
\begin{tabular}{|P{2.5cm}|P{2.5cm}|P{2.5cm}|}
\hline
\textbf{Absorbed Dose (Gy)} & \textbf{Number of Voxels} & \textbf{Voxel Position} \\ \hline
0 & 17 & Away \\
$6.50x10^{-9}$ & 2 & Adjacent \\
$3.25x10^{-8}$ & 5 & Adjacent \\
$5.85x10^{-8}$ & 8 & Adjacent \\
$1.82x10^{-7}$ & 5 & Adjacent \\
$6.63x10^{-7}$ & 1 & Adjacent \\
$8.38x10^{-7}$ & 1 & Central \\
$8.64x10^{-7}$ & 1 & Central \\
$9.74x10^{-7}$ & 1 & Central \\
$1.06x10^{-6}$ & 1 & Central \\
$1.12x10^{-6}$ & 1 & Central \\
$1.13x10^{-6}$ & 1 & Central \\
$1.20x10^{-6}$ & 1 & Central \\
$1.22x10^{-6}$ & 1 & Central \\
$1.26x10^{-6}$ & 1 & Central \\
$1.35x10^{-6}$ & 1 & Central \\
$1.38x10^{-6}$ & 1 & Central \\
$1.46x10^{-6}$ & 1 & Central \\
$1.51x10^{-6}$ & 1 & Central \\
$1.62x10^{-6}$ & 1 & Central \\
$1.66x10^{-6}$ & 1 & Central \\\hline
\end{tabular}
\label{DosisFantoma888}
\end{table}

\begin{table}
\centering
\caption{Mean, maximum and minimum dose of a 1x1x1 $mm^3$ voxel from the $^{18}F$ source of size 4 mm x 4 mm x 4 mm.}
\begin{tabular}{|P{3.5cm}|P{3.5cm}|}
\hline
Average Dose in the central voxel & $1.43x10^{-6}$ Gy\\ \hline
$D_{max}$ & $1.66x10^{-6}$ Gy \\ \hline
$D_{min}$ & $6.67x10^{-7}$ Gy \\ \hline
\end{tabular}
\label{DosisMaxMinMed1x1x1}
\end{table}

\section*{Conclusions}
By studying the simulation conditions for interactions of fluorine-18 in a voxelized phantom of 4 x 4 x 4 $mm^3$ and using the JamesRandom simulation engine, we find that the average absorbed dose is $7.10x10^ {-7}$ Gy, with a dose of $6.33x10^{-7}$ Gy in the central voxel. The percentage error across the entire calculation is 1.22 \%, indicating precise results. Using a 2x2x2 $mm^3$ resolution yields an average absorbed dose of $6.32x10^{-7}$ Gy in the central voxel, while a 1x1x1 $mm^3$ resolution results in $1.43x10^{-6}$ Gy. This shows that higher resolutions result in increased absorbed dose values compared to the 2x2x2 $mm^3$ and 4x4x4 $mm^3$ resolutions. Additionally, for a larger water volume, it was verified that gamma dose contributes to the simulation.

By entering the absorbed dose values and making a linearly dependent calculation on activity in MATLAB using corrected brain images, the absorbed dose in different brain areas with 4x4x4 $mm^3$ resolution was determined to range between 4 and 14 mGy. In lobar regions, the absorbed dose ranges between 8 and 16 mGy, while in the central region, including the thalamus and nearby regions, the absorbed dose reached a value between 2 and 8 mGy. This result indicates that the absorbed dose in brain regions can be separated into distinct subregions, demonstrating the applicability of the Monte Carlo technique to other body regions.

Finally, after performing multiple simulations with the GATE program, the absorbed dose values in the central voxel and the processed data as a DICOM image in MATLAB, yield the results that the absorbed dose values in the brain region closely align with the values reported in MIRD Pamphlet 19 and ICRP 106 Addendum 3 when using the Monte Carlo GATE Method technique. 

The above mentioned conclusions allow for high accuracy dose calculations in patient-specific dosimetry, with possible applications at the clinical level.

\section*{Acknowledgments}
We extend our sincere gratitude to the Nuclear Medicine Facility at the Hospital Universitario de Caracas and the Instituto Nacional de Bioingeniería, Universidad Central de Venezuela for their technical support and data collection that made this research possible.

\bibliographystyle{unsrt}

\end{document}